\documentclass[pra,reprint,amsmath,amssymb,aps,preprintnumbers]{revtex4-1}
\usepackage{graphicx} 
\usepackage{datetime}
\usepackage{bm} 

\begin{document}
\title{Exciton-induced transparency in hybrid plasmonic systems}
\author{Tigran V. Shahbazyan}
\affiliation{
Department of Physics, Jackson State University, Jackson, MS
39217 USA}


\begin{abstract} 
We present a microscopic model for exciton-induced transparency (ExIT) in a hybrid system comprised of an emitter resonantly coupled to a surface plasmon in a metal-dielectric structure. We obtain an effective optical polarizability of such a system with coupling  between the system components expressed in terms of energy transfer rates. We demonstrate that, in the weak coupling regime, the underlying mechanism of ExIT is the  energy exchange imbalance  between the plasmon and the emitter in a narrow frequency region. We derive in analytic form a frequency-dependent function that accurately describes the shape and amplitude of the transparency window in scattering spectra, supported by numerical calculations.
\end{abstract}
\maketitle


%
\section{Introduction}
\label{sec:intro}

Strong coupling between surface plasmons in metal-dielectric structures  and excitons in semiconductors or dye molecules has recently attracted intense interest driven to a large extent by  possible applications in ultrafast reversible switching \cite{ebbesen-prl11,bachelot-nl13,zheng-nl16}, quantum computing \cite{waks-nnano16,senellart-nnano17}, and light harvesting \cite{leggett-nl16}. In the strong coupling regime, coherent energy exchange between excitons and plasmons  \cite{shahbazyan-nl19} leads to the emergence of mixed polaritonic states with energy bands separated by an anticrossing gap (Rabi splitting) \cite{novotny-book}. For excitons coupled to  cavity modes in  microcavities, the Rabi splitting magnitudes  are  relatively small on the scale of several meV  \cite{forchel-nature04,khitrova-nphys06,imamoglu-nature06}. However,   in hybrid plasmonic systems, in which surface plasmons  are coupled to excitons in J-aggregates \cite{bellessa-prl04,sugawara-prl06,wurtz-nl07,fofang-nl08,guebrou-prl12,bellessa-prb09,schlather-nl13,lienau-acsnano14,shegai-prl15,shegai-nl17,shegai-acsphot19}, in various dye molecules \cite{hakala-prl09,berrier-acsnano11,salomon-prl12,luca-apl14,noginov-oe16},  or in semiconductor nanostructures \cite{vasa-prl08,gomez-nl10,gomez-jpcb13,manjavacas-nl11,lawrie-nl12}, the  Rabi splittings can be much greater, even reaching  hundreds  meV.

While  Rabi splitting in the emission spectra signals on strong exciton-plasmon coupling, a narrow minimum may also appear in the scattering (or absorption) spectra prior the strong coupling transition point. Such a minimum in the hybrid plasmonic system spectra has been referred to as exciton-induced transparency (ExIT)  \cite{vuckovic-prl06,bryant-prb10,pelton-oe10}, in analogy to electromagnetically induced transparency (EIT) in three-level atomic systems, and has similarly been attributed to a Fano-like interference between different excitation pathways \cite{pelton-oe10,pelton-nc18}. A recent comprehensive review \cite{pelton-ns19} has suggested that ExIT  has, in fact,  been observed  in a number of  experiments involving single excitons in J-aggregates or colloidal QDs coupled to gap plasmons in nanoparticle-on-metal (NoM) systems \cite{haran-nc16,baumberg-nature16,lienau-acsph18,pelton-nc18}, as well as in two-dimensional atomic crystals, such as WS$_2$ monolayers, conjugated with Ag or Au nanostructures \cite{shegai-acsph18,xu-nl17,alu-oe18,urbaszek-nc18,shegai-nl18,zhang-acsphot19}. Notably, for single excitons, achieving  a strong exciton-plasmon coupling is a  challenging task as it requires extremely large plasmon local density of states (LDOS) at the exciton position that can mainly be achieved in nanogaps \cite{hecht-sci-adv19,pelton-sci-adv19,baumberg-natmat2019}.

The analogy between ExIT and EIT hinges on the observation that, due to large difference (several orders) in the plasmon and exciton dipole moments, an exciton near a plasmonic structure can be viewed as a dark state that is mainly excited indirectly by the plasmon near field \cite{pelton-oe10,pelton-nc18}. The latter is thought to play the role of pump field in EIT and, at the same time, to provide the coupling between  bright (plasmon) and dark (exciton) states. Similar to EIT, the ExIT in a hybrid plasmonic system has been described by a classical model of two coupled oscillators, only one  of them interacting with the radiation field \cite{pelton-oe10}. Note, however, that, with an increasing  coupling, the system undergoes a transition to strong coupling regime, which is  characterized by coherent energy exchange  between the system components \cite{shahbazyan-nl19}. This, in turn, raises a question on whether a similar energy exchange mechanism, rather than the  Fano interference, underpins the ExIT  as well.

In this paper, we present a microscopic model describing  ExIT for a single emitter resonantly coupled to a plasmon mode in a metal-dielectric structure. We derive the effective polarizability of a hybrid plasmonic system that includes  the coupling between an exciton and plasmon  expressed in terms of the energy transfer (ET) rate between them. We elucidate the underlying ExIT mechanism by analyzing this effective polarizability in terms of both the interference of excitation pathways and the energy exchange between the system components. We show that, if the plasmon spectral linewidth is larger than that of the emitter, the  back and forth energy transfer rates are \textit{not} balanced in a narrow frequency interval, despite being equal in the entire spectral range, and that such an imbalance leads to a minimum in the scattering spectra on top of the plasmon resonance peak. We derive in analytic form the characteristic frequency-dependent function that describes the ExIT window spectral shape and amplitude. We illustrate our model by numerical calculations for an emitter near the  tip of a gold nanorod.

\section{Quantum emitter  coupled to a resonant plasmon mode}

\subsection{Optical polarizability of a plasmonic stucture}

We consider  a metal-dielectric structure characterized by a complex  dielectric function $\varepsilon (\omega,\bm{r})=\varepsilon' (\omega,\bm{r})+i\varepsilon'' (\omega,\bm{r})$ that supports localized plasmon modes with frequencies $\omega_{m}$. For characteristic system size smaller than the radiation wavelength, the plasmon modes are determined by  Gauss's law \cite{stockman-review}
%
\begin{equation}
\label{gauss}
\bm{\nabla}\!\cdot\! \left [\varepsilon' (\omega_{m},\bm{r}) \bm{\nabla}\Phi_{m}(\bm{r})\right ]=0,
\end{equation}
where 
$\Phi_{m}(\bm{r})$ is the mode potential that defines the mode field $\bm{E}_{m}(\bm{r})=-\bm{\nabla}\Phi_{m}(\bm{r})$, which we choose to be real. To determine the plasmon dipole moment for optical transitions, we recast Eq.~(\ref{gauss}) as $\bm{\nabla}\!\cdot\! \left [\bm{E}_{m}(\bm{r})+4\pi \bm{P}_{m}(\bm{r})\right ]=0$,  where $\bm{P}_{m}(\bm{r})=\chi'(\omega_{m},\bm{r})\bm{E}_{m}(\bm{r})$ is  the electric polarization vector and $\chi=(\varepsilon-1)/4\pi$  is the plasmonic system susceptibility. The plasmon  dipole moment has the form
%
\begin{equation}
\bm{p}_{m}=\int dV \bm{P}_{m}=\int dV \chi'(\omega_{m},\bm{r})\bm{E}_{m}(\bm{r}).
\end{equation}
Although Gauss's equation (\ref{gauss}) does not, by itself,  determine the overall field normalization \cite{stockman-review}, the latter can be set, e.g., by matching the plasmon radiative decay rate to that of a localized dipole with excitation energy $\hbar \omega_{m}$. The radiative decay rate of a plasmon mode has a standard form \cite{shahbazyan-prb18} $\gamma_{m}^{r}=W_{m}^{r}/U_{m}$, where 
\begin{align}
\label{energy-mode}
U_{m}
= \frac{1}{16\pi} 
\!\int \!  dV \, \frac{\partial [\omega_{m}\varepsilon'(\omega_{m},\bm{r})]}{\partial \omega_{m}} \,\bm{E}_{m}^{2}(\bm{r}) 
\end{align}
is the plasmon mode energy \cite{landau,shahbazyan-prl16} and 
\begin{equation}
\label{power}
W_{m}^{r}=\frac{p_{m}^{2}\omega_{m}^{4}}{3c^{3}}
\end{equation}
is the radiated power \cite{novotny-book}. The normalized modes $\tilde{\bm{E}}_{m}(\bm{r})$ are determined by setting $\gamma_{m}^{r}=4\mu_{m}^{2}\omega_{m}^{3}/3\hbar c^{3}$, where $c$ is the speed of light and $\mu_{m}$ is the mode optical transition matrix element. We then find the relation
\begin{equation}
\label{mode-rescaled}
\tilde{\bm{E}}_{m}(\bm{r})=\frac{1}{2}\sqrt{\frac{\hbar\omega_{m}}{U_{m}}} \bm{E}_{m}(\bm{r}),
\end{equation}
and, accordingly, $\bm{\mu}_{m}=\int dV \chi'(\omega_{m},\bm{r})\tilde{\bm{E}}_{m}(\bm{r})$ (the factor $1/2$ reflects positive-frequency contribution).  Hereafter, we  will  use  the normalized modes, unless noted.

The response of plasmonic structure to an external field $\bm{E}_{in}e^{-i\omega t}$ is characterized by the polarizability tensor \cite{shahbazyan-prb18} $\bm{\alpha}_{pl}=\sum_{m}\bm{\alpha}_{m}$, where $\bm{\alpha}_{m}(\omega)=\alpha_{m}(\omega)\bm{n}_{m}\bm{n}_{m}$ is the  mode polarizability tensor ($\bm{n}_{m}$ is the plasmon mode polarization). Near the resonance, keeping only the resonant term, the mode scalar polarizability has the form 
\begin{equation}
\label{polar-mode}
\alpha_{m}(\omega)=\frac{\mu_{m}^{2}}{\hbar}\frac{1}{\omega_{m}-\omega-\frac{i}{2}\gamma_{m}},
\end{equation}
where, $\gamma_{m} =\gamma_{m}^{nr}+\gamma_{m}^{r}$ is the plasmon decay rate that is comprised of  radiative rate $\gamma_{m}^{r}$ and  nonradiative rate  $\gamma_{m}^{nr}=2\varepsilon''(\omega_{m})/[\partial\varepsilon'(\omega_{m})/\partial\omega_{m}]$ due to Ohmic losses.  
The scattering cross-section of a plasmon mode is given by a standard relation  $\sigma_{m}^{sc}(\omega)\propto \omega^{4}|\alpha_{m}(\omega)|^{2} $ and, near the resonance, has a simple form  
\begin{equation}
\label{mode-sc}
\sigma_{m}^{sc}(\omega)\propto 
\frac{\omega^{4}}{(\omega_{m}-\omega)^{2}+\gamma_{m}^{2}/4},
\end{equation}
where we omitted a constant prefactor.

\subsection{Effective optical polarizability of a hybrid plasmonic system}

Let us now consider a quantum emitter (QE) situated at  a position $\bm{r}_{e}$ near a metal-dielectric structure. The optical response of   a hybrid system can be described in terms of \textit{effective polarizability} $\bm{\alpha}_{s}(\omega)$ that includes QE-plasmon optical interactions. Here we consider the weak coupling regime and treat plasmons classically. Typically, the QE optical transition matrix element   $\bm{\mu}_{e}=\mu_{e} \bm{n}_{e}$, where $\bm{n}_{e}$ is the dipole orientation,  is much smaller (by several orders) than $\bm{\mu}_{m}$  and, therefore, direct QE interaction with the radiation field can be neglected \cite{pelton-oe10,pelton-nc18}. Instead, the QEs are excited indirectly by the local field $\tilde{\bm{E}}_{m}(\bm{r}_{e})$  of resonantly excited  plasmon mode. The plasmon-induced QE dipole moment has the standard form 
\begin{equation}
\bm{p}_{e}=\bm{\alpha}_{e}(\omega)\tilde{\bm{E}}_{m}(\bm{r}_{e}),
\end{equation}
where  $\bm{\alpha}_{e}(\omega)=\alpha_{e}(\omega)\bm{n}_{e}\bm{n}_{e}$ is the QE optical polarizability tensor. Since QE excitation is a secondary effect, the effective polarizability of a hybrid system can be obtained, within the dressed plasmon picture,  by appropriately modifying the plasmon polarizability (\ref{polar-mode}). Namely, the back-interaction of the plasmon-induced QE dipole with the plasmon   is described by plasmon self-energy 
\begin{equation}
\label{self}
\hbar \Sigma_{m}(\omega)=- \bm{p}_{e}\!\cdot\! \tilde{\bm{E}}_{m}(\bm{r}_{e})
=-\alpha_{e}(\omega)[\bm{n}_{e}\!\cdot\! \tilde{\bm{E}}_{m}(\bm{r}_{e})]^{2},
\end{equation}
which should be added to the plasmon energy $\hbar\omega_{m}$. 

The effective polarizability tensor of a  hybrid system near the resonance takes the form
\begin{equation}
\label{polar-eff}
\alpha_{s}(\omega)=\frac{\mu_{m}^{2}}{\hbar}\frac{1}{\omega_{m}+\Sigma_{m}(\omega)-\omega-\frac{i}{2}\gamma_{m}}.
\end{equation}
We assume that the QE excitation frequency $\omega_{e}$ is close to the plasmon frequency $\omega_{m}$ and, therefore, near the resonance, adopt a classical QE polarizability   \cite{novotny-book}
\begin{equation}
\label{polar-qe}
 \alpha_{e}(\omega)=\frac{\mu_{e}^{2}}{\hbar}\frac{2\omega_{e}}{\omega_{e}^{2}-\omega^{2}-i\omega\gamma_{e}}
 \approx
 \frac{\mu_{e}^{2}}{\hbar}\frac{1}{\omega_{e}-\omega-\frac{i}{2}\gamma_{e}},
 \end{equation} 
where $\gamma_{e}$ is the QE spectral linewidth assumed here to be much smaller than the plasmon one, $\gamma_{e}\ll \gamma_{m}$. Using Eq.~(\ref{polar-qe}), the plasmon self-energy (\ref{self}) takes the form
\begin{equation}
\label{self-coupling}
\Sigma_{m}(\omega)=-\frac{g^{2}}{\omega_{e}-\omega-\frac{i}{2}\gamma_{e}},
\end{equation}
where $\hbar g=-\bm{\mu}_{e}\!\cdot\! \tilde{\bm{E}}_{m}(\bm{r}_{e})$ is the QE-plasmon coupling parameter. Returning, for a moment, to the original (not normalized) plasmon mode fields (\ref{mode-rescaled}), we recover a cavitylike expression \cite{shahbazyan-nl19}
\begin{equation}
\label{qe-pl-coupling-mode-volume}
g^{2} = \frac{2\pi \mu_{e}^{2}\omega_{m} }{\hbar {\cal V}},
~~~
\frac{1}{{\cal V}}  = \frac{2[\bm{n}_{e}\!\cdot\! \bm{E}_{m}(\bm{r}_{e})]^{2}}{\int \! dV [\partial (\omega_{m}\varepsilon')/\partial \omega_{m}]\bm{E}_{m}^{2}},
\end{equation}
where ${\cal V}$ is the projected plasmon mode volume that characterizes  the plasmon field confinement at the QE position $\bm{r}_{e}$ along its dipole orientation $\bm{n}_{e}$ \cite{shahbazyan-prl16,shahbazyan-acsphot17,shahbazyan-prb18}. 

The effective polarizability of a hybrid system near the resonance is   obtained by inserting Eq.~(\ref{self-coupling}) into Eq.~(\ref{polar-eff}):
%
\begin{equation}
\label{polar-full2}
 \alpha_{s}(\omega)
=
\frac{\mu_{m}^{2}}{\hbar}
\frac{ \omega_{e}-\omega - \frac{i}{2}\gamma_{e}}
{\left (\omega_{m}-\omega -\frac{i}{2}\gamma_{m}\right )\!\!\left (\omega_{e}-\omega -\frac{i}{2}\gamma_{e}\right )-g^{2}}.
\end{equation}
Thus, for a QE decoupled from the radiation field, the effective polarizability (\ref{polar-full2}) is similar to that for two coupled oscillators \cite{pelton-oe10}, but with  exciton-plasmon coupling $g$   now expressed in terms of the plasmon mode volume, as given by Eq.~(\ref{qe-pl-coupling-mode-volume}).

\subsection{Emitter-plasmon coupling and energy transfer}

The above QE-plasmon coupling $g$ can be related to the corresponding QE-plasmon ET rate \cite{shahbazyan-nl19}. Namely, the rate $\gamma_{e\rightarrow m}(\omega)$ for transferring energy $\hbar\omega$  from a QE to a plasmon is given by the Fermi golden rule as
\begin{equation}
\label{rate-fermi}
\gamma_{e\rightarrow m}(\omega)=\frac{2\pi}{\hbar}\left |\bm{\mu}_{e}\!\cdot\! \tilde{\bm{E}}_{m}(\bm{r}_{e})\right |^{2}f_{m}(\omega),
\end{equation}
where 
\begin{equation}
\label{mode-spectrum}
f_{m}(\omega)=\frac{1}{2\pi\hbar}\frac{\gamma_{m}}{(\omega-\omega_{m})^{2}+\gamma_{m}^{2}/4}
\end{equation}
is  the plasmon spectral function satisfying $\hbar\!\int\! d\omega f_{m}(\omega)=1$. Using the relation $ g=-\bm{\mu}_{e}\!\cdot\! \tilde{\bm{E}}_{m}(\bm{r}_{e})/\hbar$, the frequency-resolved QE-plasmon ET rate (\ref{rate-fermi}) takes the form 
\begin{equation}
\label{rate-em}
\gamma_{e\rightarrow m}(\omega)=\frac{g^{2}\gamma_{m}}{(\omega-\omega_{m})^{2}+\gamma_{m}^{2}/4}.
\end{equation}
At resonance ($\omega=\omega_{m}$), we obtain an important relation 
\begin{equation}
\label{coupling-rate}
g^{2}=\frac{1}{4}\gamma_{m}\gamma_{e\rightarrow m},
\end{equation}
where hereafter we use the notations   $\gamma_{e\rightarrow m}\equiv \gamma_{e\rightarrow m}(\omega_{m})$. Comparing to Eq.~(\ref{qe-pl-coupling-mode-volume}), the QE-plasmon ET rate is expressed via the plasmon mode volume as
\begin{equation}
\label{qe-plasmon-rate-mode-volume}
\gamma_{e\rightarrow m}= \frac{8\pi \mu_{e}^{2}Q_{m} }{\hbar {\cal V}},
\end{equation}
where $Q_{m} =\omega_{m}/\gamma_{m}$ is the plasmon quality factor. Recalling that the Purcell factor is  $F_{p}=\gamma_{e\rightarrow m}/\gamma_{e}^{r}$, where $\gamma_{e}^{r}=4\mu_{e}^{2}\omega^{3}/3\hbar c^{3}$ is the QE radiative decay rate, we recover the cavitylike expression for the Purcell factor in terms of the plasmon mode volume: $F_{p}=6\pi c^{3}Q_{m}/\omega^{3} {\cal V}$.

\section{Exciton-induced transparency}


\subsection{Excitation pathways interference picture}

The effective  polarizability  Eq.~(\ref{polar-full2}) possesses two resonances in the complex frequency plane assigned to  polaritonic bands  $\omega_{\pm}=\frac{1}{2}[\omega'_{m}+\omega'_{e}\pm \sqrt{(\omega'_{m}-\omega'_{e})^{2}+4g^{2}}]$, where $\omega'_{m}=\omega_{m}-i\gamma_{m}/2$ and $\omega'_{e}=\omega_{e}-i\gamma_{e}/2$.  Assume, for a moment,  that  QE and plasmon frequencies are in resonance ($\omega_{e}=\omega_{m}$). In the weak coupling regime, the polaritonic bands are energy degenerate but have  different linewidths \cite{pelton-oe10},
\begin{equation}
\omega_{\pm}=\omega_{m}-\frac{i}{2}\left [\frac{\gamma_{m}+\gamma_{e}}{2}\pm \sqrt{(\gamma_{m}-\gamma_{e})^{2}/4-4g^{2}}\right ],
\end{equation}
where it is assumed that the square root is positive. In terms of transitions to  polaritonic states, the effective polarizability Eq.~(\ref{polar-full2}) can be presented as
\begin{align}
\label{polar-so}
\alpha_{s}(\omega)
= \frac{\mu_{m}^{2}}{\hbar}
\left ( \frac{1+a}{\omega_{+}-\omega}+\frac{1-a}{\omega_{-}-\omega}\right ),
\end{align}
where the parameter $a$ is given by
\begin{equation}
a=\frac{(\gamma_{m}-\gamma_{e})/2}{\sqrt{(\gamma_{m}-\gamma_{e})^{2}/4-4g^{2}}}.
\end{equation}
Note that, even at resonance, the interference between two excitation pathways is neither purely constructive nor destructive but, in fact, is the admixture of both   controlled by the  parameter $a$. In the absence of QE-plasmon coupling (i.e., $a=1$), the effective polarizability (\ref{polar-so}) reduces to the plasmon polarizability (\ref{polar-mode}) reflecting the fact that the QE is not coupled to the radiation field. With the QE-plasmon coupling $g$ turned on,  the parameter $a$ increases up until the strong coupling transition point, at which it becomes imaginary. Since $a>1$ prior the the transition, the system absorption spectrum, described by $\alpha''_{s}(\omega)$, exhibits a narrow minimum (ExIT). However, this minimum has no specific onset and, therefore, does not imply a  separate (intermediate) phase, in contrast to the strong coupling regime characterized by a clear transition point. At   resonance frequency ($\omega=\omega_{m}=\omega_{e}$), the system effective polarizability is purely imaginary $\alpha_{s} \propto i(\gamma_{m}+4g^{2}/\gamma_{e})^{-1}$. Normalizing $\alpha_{s}$  by the plasmon polarizability (\ref{polar-mode}) at resonance frequency, $\alpha_{m}\propto i/\gamma_{m}$, and using the relation (\ref{coupling-rate}) between the QE-plasmon coupling and ET rate, we obtain the ratio of  the absorption spectra,  at resonance frequency, for the hybrid system and plasmon mode,
\begin{equation}
\label{exit-minimum}
\frac{\alpha''_{s} }{\alpha''_{m} }=\frac{\gamma_{e}}{\gamma_{e}+\gamma_{e\rightarrow m}},
\end{equation}
which characterizes the ExIT minimum depth. With increasing QE-plasmon ET rate $\gamma_{e\rightarrow m}$, the ratio (\ref{exit-minimum}) steadily decreases crossing over to the strong coupling regime ($|a|=\infty$), where the ExIT minimum turns into the Rabi splitting. Importantly, Eq.~(\ref{exit-minimum}) is independent of the parameter $a$, which controls the interference between excitation pathways, and therefore is \textit{not} sensitive to the transition point. This suggests an  interpretation of ExIT in terms of QE-plasmon energy exchange  that governs the strong coupling regime as well.

\subsection{Energy exchange picture}

In the steady state, as the system is continuously illuminated by monochromatic light, the \textit{full} back and forth ET rates between a QE and a plasmon should coincide.  However, for $\gamma_{e}\ll \gamma_{m}$,  the ET balance can be violated in a narrow frequency interval, leading to  distinct spectral features. To demonstrate this effect, we note that  the frequency-resolved QE-plasmon ET rate (\ref{rate-em}) is proportional, as it should be \cite{novotny-book}, to the acceptor (i.e., plasmon) absorption spectrum  $\alpha''_{m}(\omega)$. At the same time, the \textit{reverse}  plasmon-QE rate is related to the plasmon self-energy (\ref{self-coupling}) as $\gamma_{m\rightarrow e}(\omega)=-2\Sigma''_{m}(\omega)$, or
\begin{equation}
\label{rate-me}
\gamma_{m\rightarrow e}(\omega)= \frac{ g^{2} \gamma_{e}}{(\omega-\omega_{e})^{2}+\gamma_{e}^{2}/4},
\end{equation}
which  is also determined by the acceptor (QE) absorption spectrum $\alpha''_{e}(\omega)$ [see Eq.~(\ref{polar-qe})]. To obtain the full plasmon-QE ET rate  $\Gamma_{m\rightarrow e}$,  the frequency-resolved rate $\gamma_{m\rightarrow e}(\omega)$ should be integrated  \cite{novotny-book} with the normalized plasmon spectral function $f_{m}(\omega)$, given by Eq.~(\ref{mode-spectrum}): 
\begin{equation}
\Gamma_{m\rightarrow e}=\int d\omega f_{m}(\omega) \gamma_{m\rightarrow e}(\omega).
\end{equation}
Similarly, the full QE-plasmon ET rate  $\Gamma_{e\rightarrow m}$ is obtained by integrating the corresponding frequency-resolved rate (\ref{rate-em}) with the analogous normalized QE spectral function.   Both rates are easily evaluated and we obtain
\begin{equation}
\Gamma_{m\rightarrow e}=\Gamma_{e\rightarrow m}=\frac{g^{2}(\gamma_{m}+\gamma_{e}) }{(\omega_{m}-\omega_{e})^{2}+(\gamma_{m}+\gamma_{e})^{2}/4},
\end{equation}
indicating an overall energy exchange balance.

Near the resonance, however, the frequency-resolved plasmon-QE ET rate Eq.~(\ref{rate-me}) can be much faster than the QE-plasmon ET rate Eq.~(\ref{rate-em}) due to a  sharper QE absorption peak. Namely, for $\omega=\omega_{m}=\omega_{e}$, we have 
%
\begin{equation}
\label{rates-imbalance}
\frac{\gamma_{m\rightarrow e}}{\gamma_{e\rightarrow m}}= \frac{\gamma_{m}}{\gamma_{e}}\gg 1,
\end{equation}
implying  significant ET  excess in a narrow frequency interval, to be  compensated at frequencies beyond  this interval. Such imbalance between  near-resonance ET rates leads to a  narrow minimum in the dressed plasmon spectrum  as the states of a hybrid system are redistributed between its interacting components. Since the incident light mainly couples to the plasmon,  a QE-induced minimum in the dressed plasmon spectral band results in an enhanced light transmission (ExIT).
 
To elucidate the emergence of ExIT minimum, we recall the relation between the system scattering cross-section   $\sigma_{s}^{sc}$ and its polarizability:  $\sigma_{s}^{sc}(\omega)\propto \omega^{4}|\alpha_{s}(\omega)|^{2} $. Comparing Eqs.~(\ref{polar-mode}) and (\ref{polar-full2}), we obtain
\begin{equation}
\label{sigma-sc}
\sigma_{s}^{sc}(\omega)=\sigma_{m}^{sc}(\omega)R(\omega),
\end{equation}
where $\sigma_{m}^{sc}(\omega)$ is given by Eq.~(\ref{mode-sc}) and  
\begin{equation}
\label{R-full}
R(\omega)=\left |\frac{\left (\omega_{m}-\omega-\frac{i}{2}\gamma_{m}\right )\left (\omega_{e}-\omega-\frac{i}{2}\gamma_{e}\right )}
{\left (\omega_{m}-\omega-\frac{i}{2}\gamma_{m}\right )\left (\omega_{e}-\omega-\frac{i}{2}\gamma_{e}\right )-g^{2}}\right |^{2}
\end{equation}
is a frequency-dependent function that, in the weak coupling regime,   modulates the plasmon cross-section $\sigma_{m}^{sc}(\omega)$. In a  narrow frequency interval $|\omega_{m}-\omega|/\gamma_{m}\ll 1$, using the relation (\ref{coupling-rate}), the function  $R(\omega)$  simplifies to
\begin{equation}
\label{R-weak}
E(\omega)=\frac{\epsilon^{2}+1}
{\epsilon^{2}+(1+p)^{2}},
\end{equation}
where $\epsilon=2(\omega-\omega_{e})/\gamma_{e}$ is  the detuning from the emitter frequency normalized by its linewidth, and  
\begin{equation}
\label{exitP}
p=\frac{\gamma_{e\rightarrow m}}{\gamma_{e}}=\frac{\gamma_{m\rightarrow e}}{\gamma_{m}}
\end{equation}
is the parameter characterizing the  ExIT minimum depth [compare to Eq.~(\ref{exit-minimum})]. 

We observe that the ExIT function $E(\omega)$ is distinct from the Fano function $F(\omega)=(\epsilon-q)^{2}/(\epsilon^{2}+1)$, which arises from  the interference between a localized state and continuum. Indeed, the Fano  parameter $q$ defines the frequency, away from the resonance, at which the destructive interference occurs, whereas the ExIT parameter $p$ modifies the plasmon decay rate near the emitter frequency. Namely, in the weak coupling regime, the decay rate of a dressed plasmon resonantly coupled to a QE has the form $\gamma_{m}^{s}(\omega)=\gamma_{m}+\gamma_{m\rightarrow e}(\omega)$.  Using Eq.~(\ref{rate-me}) and the relation (\ref{coupling-rate}), we obtain
\begin{equation}
\label{plasmon-rate-qe}
\gamma_{m}^{s}(\omega)
=\gamma_{m}\left (1+ \frac{ p}{\epsilon^{2}+1}\right ),
\end{equation}
indicating that the dressed plasmon linewidth increases by  factor $(1+p)$  in  the frequency interval $|\omega-\omega_{e}|\sim \gamma_{e}$. Since a linewidth increase is accompanied by  amplitude drop, this   leads to a dip in the dressed plasmon spectrum in that frequency interval. 

\section{Discussion and numerical results}

%
\begin{figure}[b]
\begin{center}
\includegraphics[width=0.99\columnwidth]{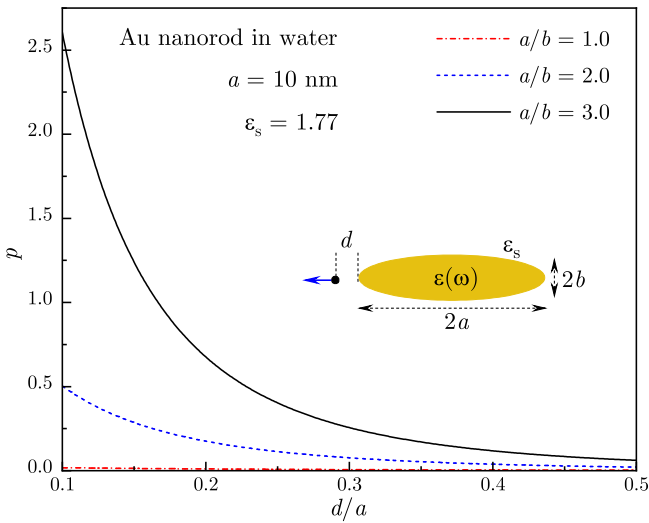}
\caption{\label{fig1} ExIT parameter $p$ for an emitter near a tip of gold nanorod in water is plotted against the distance to the tip for several values of nanorod aspect ratio. Inset: schematics of an emitter near  Au nanorod tip.
 }
\end{center}
\vspace{-3mm}
\end{figure}
%

Below we present the results of numerical calculations for an emitter situated at a distance $d$ from the tip of an Au nanorod in water modeled by a prolate spheroid with semimajor and semiminor axes $a$ and $b$, respectively (see Fig.~\ref{fig1}). The emitter's dipole orientation is chosen along the nanorod symmetry axis, the nanorod overall size is $2a=20$ nm, standard spherical harmonics were used for modeling the longitudinal plasmon fields, and the Au experimental dielectric function  is used in all calculations \cite{johnson-christy}. The emitter spectral linewidth $\gamma_{e}$ is much smaller than that of plasmon, $\gamma_{e}/\gamma_{m}=0.1$, while its radiative decay time is chosen $\tau_{e}^{r}= 10$ ns, which are typical values for excitons in semiconductor quantum dots. Note that the  emitter's radiative decay rate $\gamma_{e}^{r}$ is much smaller that its spectral linewidth: for our system we have $\gamma_{e}^{r}/\gamma_{e}\sim 10^{-5}$. For such values, the transition to strong coupling regime for a single emitter requires extremely large Purcell factors that are not normally achieved for free-standing nanorods, so that all the results below are obtained in the weak coupling regime.

%
\begin{figure}[t]
\begin{center}
\includegraphics[width=0.99\columnwidth]{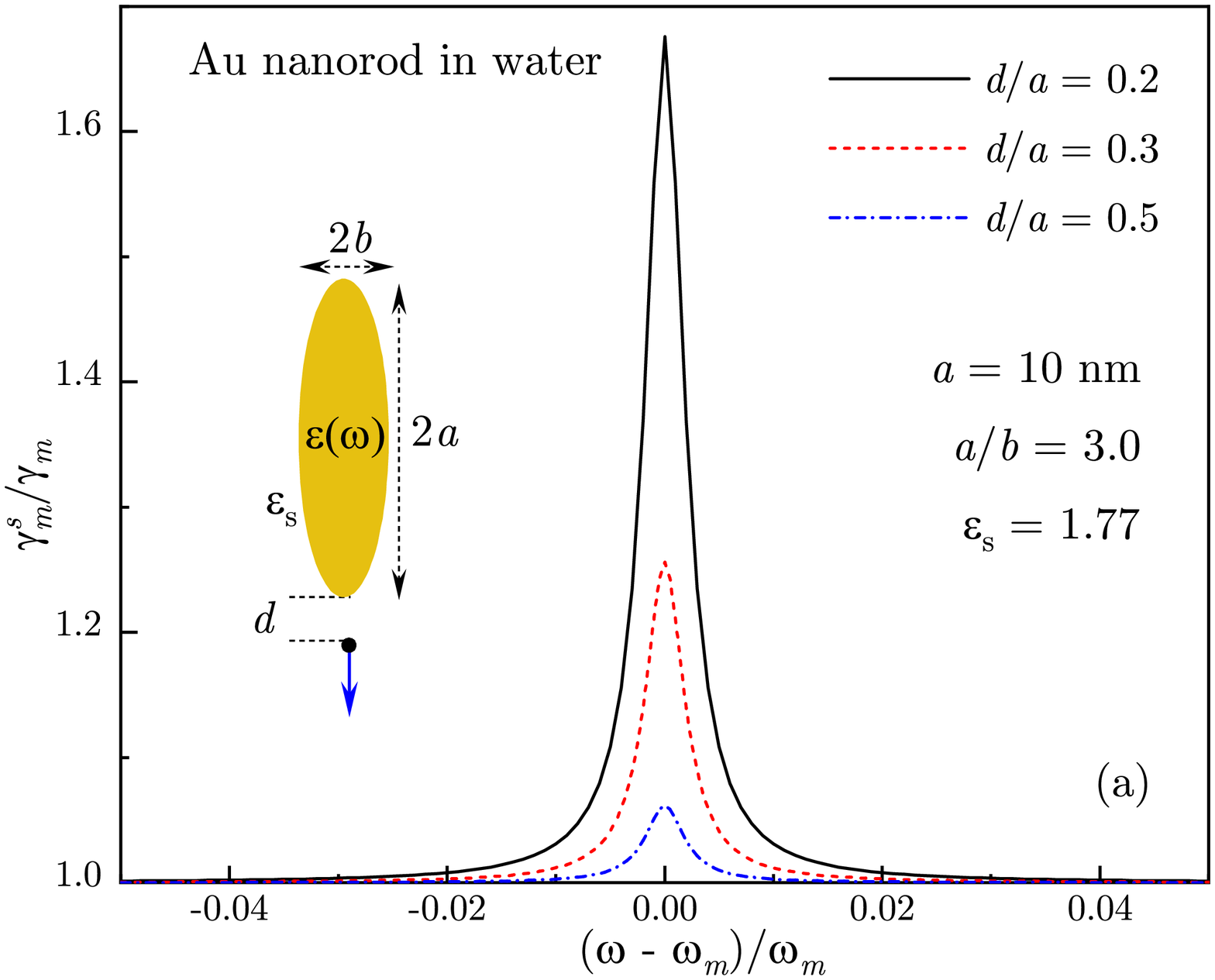}

\vspace{2mm}

\includegraphics[width=0.99\columnwidth]{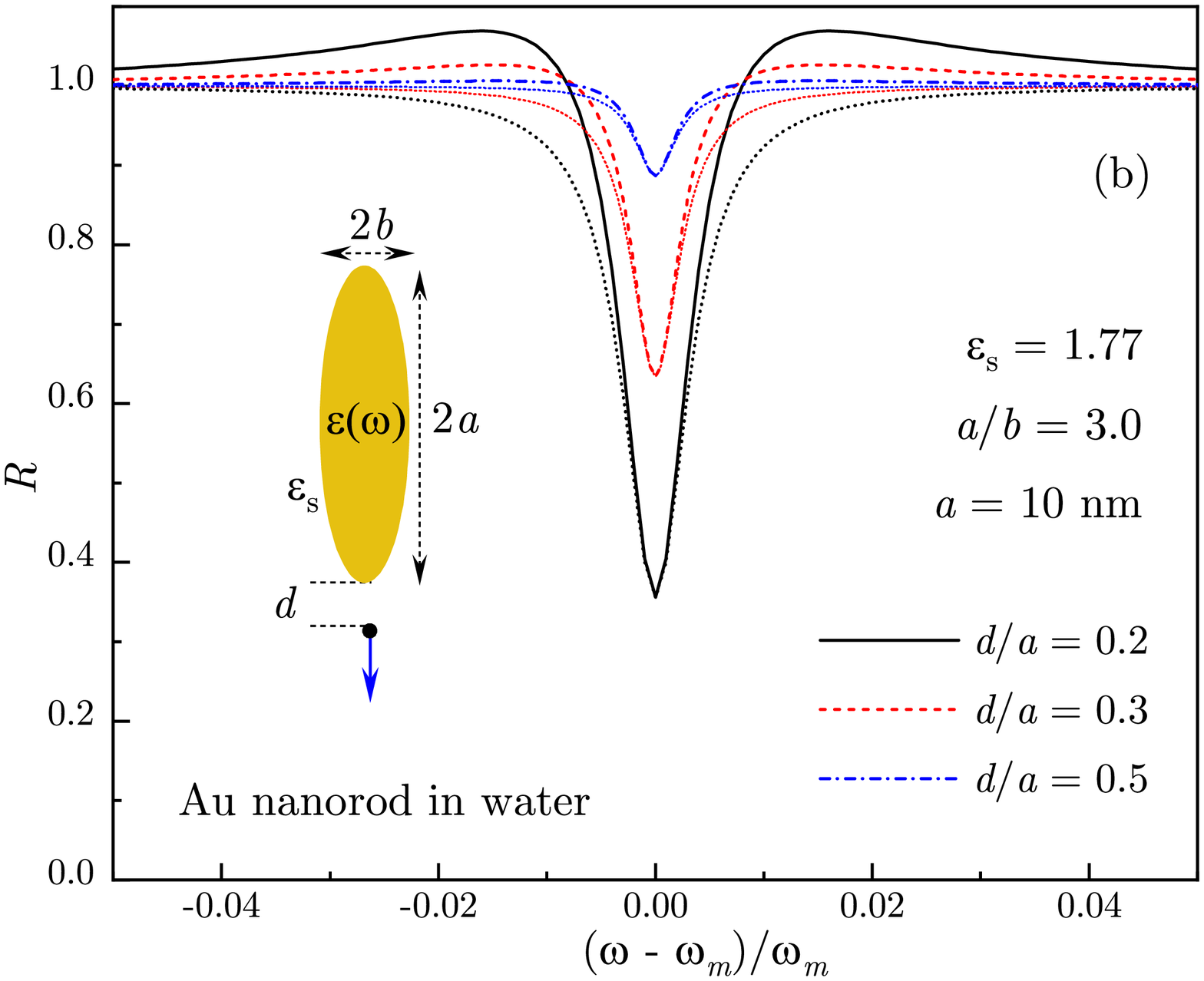}
\caption{\label{fig2} (a) Normalized  decay rate of a dressed plasmon is shown for several values of $d$. (b) System scattering cross-section  relative to the plasmon cross-section, given by Eq.~(\ref{R-full}), and the corresponding ExIT function $E(\omega)$ (dotted lines) are shown   for several values of $d$.
 }
\end{center}
\vspace{-6mm}
\end{figure}
%

In Fig.~\ref{fig1}, we plot  the ExIT parameter $p=\gamma_{e\rightarrow m}/\gamma_{e}=F_{p}\gamma_{e}^{r}/\gamma_{e}$ against the emitter's distance to nanorod tip for several values of aspect ratio $a/b$. Note that the Purcell factor is largest near the tip of elongated particles, where the plasmon mode volume is small, so that $p\sim 1$ for the nanorod with  aspect ratio   ($a/b=3$), but it is  negligibly small for a nanosphere ($a/b=1$). Away from the tip,  $p$ drops sharply to $p<0.1$ at $d=0.5a$. However, even for small values of $p$, the dressed plasmon's decay rate Eq.~(\ref{plasmon-rate-qe}) still shows a spike at the emitter's frequency, which develops into a pronounced peak with reducing $d$ [see  Fig.~\ref{fig2}(a)]. This rise of  the dressed plasmon decay rate in a narrow frequency region originates from the difference between QE-plasmon and plasmon-QE (back and forth) ET rates in that region [see Eq.~(\ref{rates-imbalance})]. The same effect defines the shape of function $R(\omega)$, given by Eq.~(\ref{R-full}), which modulates the plasmon spectrum [see  Fig.~\ref{fig2}(b)]. In order to signify the role of ExIT parameter $p$,  we also plot the ExIT function  $E(\omega)$, given by Eq.~(\ref{R-weak}), for each value of QE-nanorod distance $d$ (dotted lines). Clearly, deep in the weak coupling regime (small $p$), the ExIT function $E(\omega)$ accurately describes the spectral minimum (blue curves), while for larger $p$ (i.e., closer to the tip) the spectrum develops "wings" outside the dip region as the system approaches the strong coupling transition point. Importantly, for \textit{any} distance $d$, the ExIT function $E(\omega)$ very accurately reproduces the central part of  ExIT minimum  and, in particular, its amplitude, implying that it is the energy exchange mechanism,  rather than a Fano-like interference, which is responsible for ExIT.

%
\begin{figure}[tb]
\begin{center}
\includegraphics[width=0.99\columnwidth]{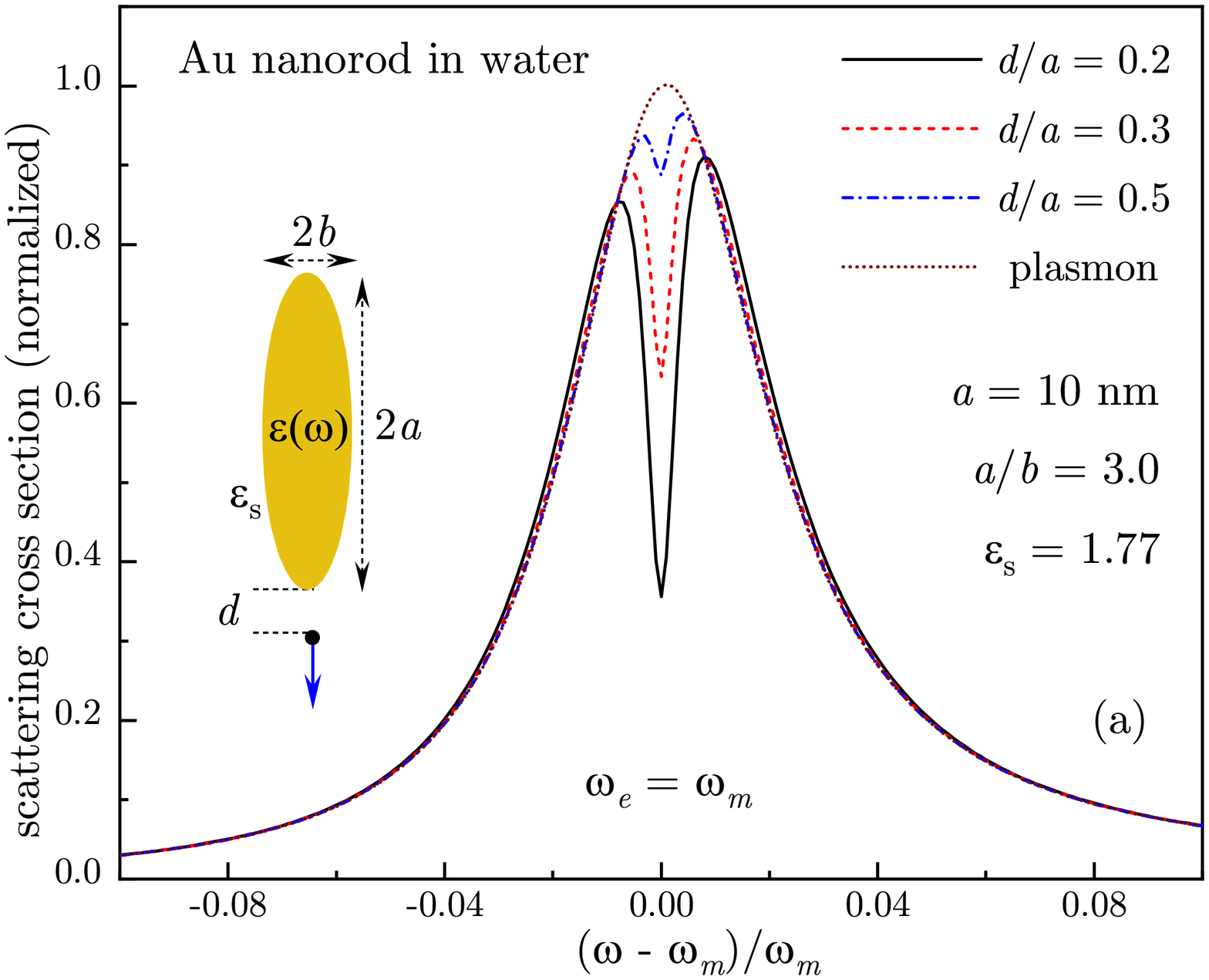}

\vspace{2mm}

\includegraphics[width=0.99\columnwidth]{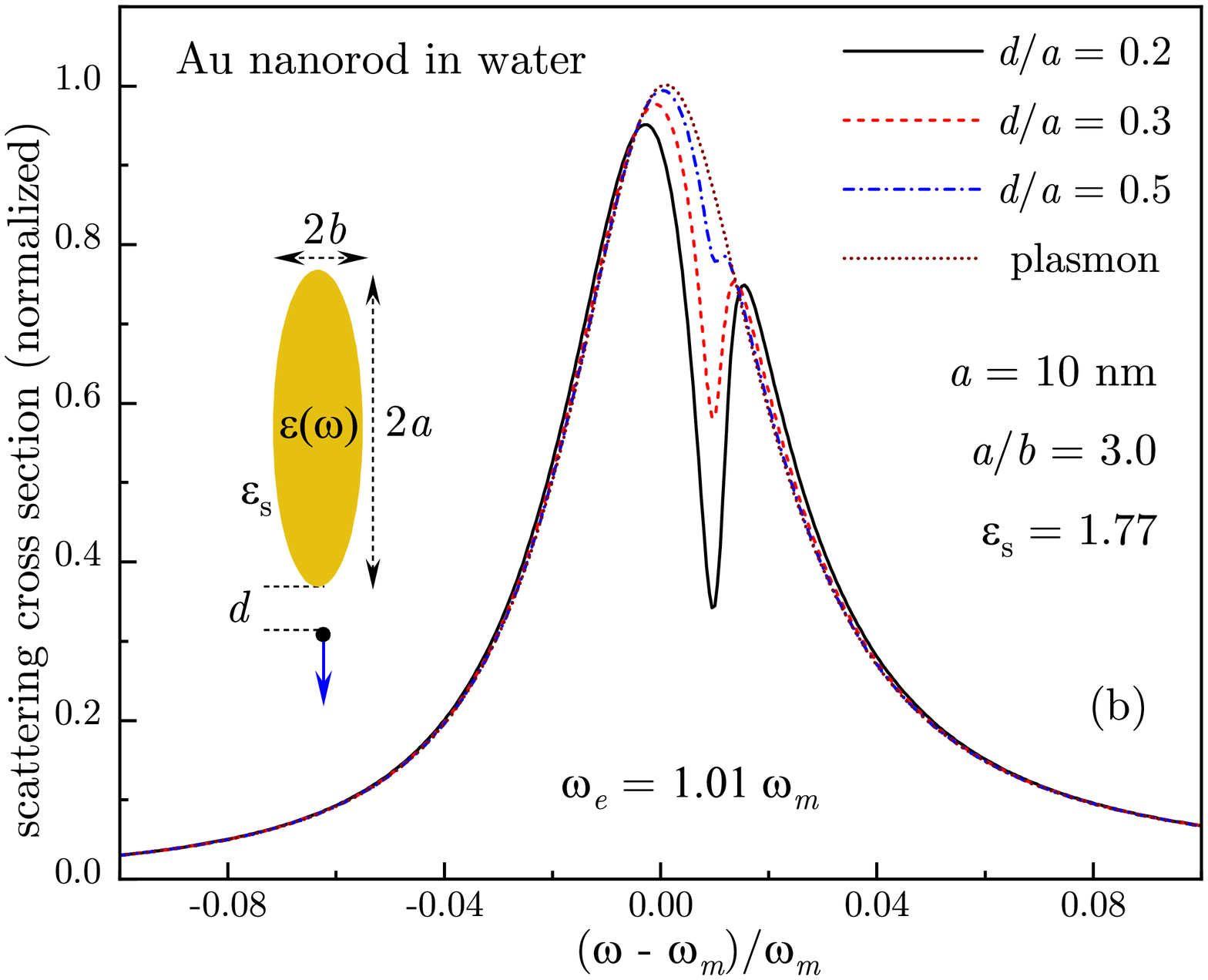}
\caption{\label{fig3} System scattering cross-section at the emitter frequency (a) in resonance with and (b) blueshifted from the plasmon frequency is shown for several values of $d$.
 }
\end{center}
\end{figure}
%

In Fig.~\ref{fig3}, we plot the normalized scattering cross-section (\ref{sigma-sc}) of the hybrid system for several values of $d$. In the weak coupling regime, the overall spectral shape is described by plasmon resonance peak modulated by the ExIT function $E(\omega)$ which exhibits a narrow minimum at the emitter's frequency $\omega_{e}$.  For  the emitter and plasmon  frequencies in exact resonance ($\omega_{e}=\omega_{m}$), the ExIT window is positioned at the center of the plasmon spectral band [see Fig.~\ref{fig2}(a)], but for $\omega_{e}$ blueshifted relative to $\omega_{m}$, the transmission maximum shifts to a higher frequency as well [see Fig.~\ref{fig2}(b)]. The fact that the ExIT window is always centered at the emitter's absorption peak position, as described by  Eq.~(\ref{R-weak}), is readily consistent with the energy exchange mechanism of ExIT but has no  natural interpretation in terms of Fano interference. Note that, even at exact resonance  ($\omega_{e}=\omega_{m}$), the double-peak spectrum is asymmetric [see Fig.~\ref{fig2}(a)] since the scattering cross-section is proportional to $\omega^{4}$ [see Eq.~(\ref{mode-sc})], reflecting the fact that, for higher frequencies,  the reemission takes place at a higher rate ($\gamma_{m}^{r}\propto \omega^{3}$). Finally, the emergence of ExIT window is not characterized by any clear onset, implying the absence of a distinct "intermediate" coupling phase.

\section{Conclusions}
In summary, we developed a model for exciton-induced transparency in hybrid plasmonic systems based on an energy exchange mechanism between the system components. For a single emitter resonantly coupled to a surface plasmon in a metal-dielectric structure, we derived an effective optical polarizability that includes exciton-plasmon coupling expressed in terms of the energy transfer rate. We analyzed in detail possible ExIT mechanisms to show that the spectral minimum in the weak coupling regime results from the energy exchange imbalance between the system components in a narrow frequency interval. We derived in analytic form a frequency-dependent function that describes accurately the shape and amplitude of the transparency window centered at the emitter absorption peak position.

\acknowledgments
This work was supported in part by National Science Foundation  Grants  No. DMR-2000170, No. DMR-1856515,  No.  DMR-1826886, and No. HRD-1547754.


\end{document}